\documentclass[12pt]{iopart}

\usepackage[colorlinks,linkcolor=blue,urlcolor=blue,citecolor=blue,pdfusetitle]{hyperref}
\usepackage[utf8]{inputenc}
\usepackage[english]{babel}
\usepackage[caption = false]{subfig}
\usepackage{graphicx,epstopdf}
\usepackage{blindtext}
\usepackage[table,xcdraw]{xcolor}
\usepackage{lipsum}
\usepackage{amsfonts}
\usepackage{bbm}
\usepackage{amssymb}
\usepackage{color}
\usepackage{latexsym}
\usepackage{times,txfonts}

\usepackage{ulem}

\newcommand{\Fcal}{\mathcal{F}}

\newcommand{\Rmath}{\mathbbm{R}}

\newcommand{\Zmath}{\mathbbm{Z}}

\newcommand{\ket}[1]{| #1 \rangle}

\newcommand{\abs}[1]{| #1 |}

\newcommand{\SubFig}[2]{\ref{#1}{\color{blue}#2}}
\newcommand{\eqref}[1]{(\ref{#1})}

\newcommand{\UFSCar}{Departamento de Física, Universidade Federal de São Carlos, Rodovia Washington Luís, km 235 - SP-310, 13565-905 São Carlos, SP, Brazil}
\newcommand{\CSIC}{Instituto de Física Fundamental, Consejo Superior de Investigaciones Científicas, Calle Serrano 113b, 28006 Madrid, Spain}

\usepackage{orcidlink}

\begin{document}
	
	\title[Scalable quantum eraser with superconducting integrated circuits]{Scalable quantum eraser with superconducting integrated circuits}

	\author{Ciro Micheletti Diniz~\orcidlink{0000-0002-7602-0468}}
	\address{\UFSCar}
	\ead{ciromd@outlook.com.br}
	
	\author{Celso J. Villas Bôas~\orcidlink{0000-0001-5622-786X}}
	\address{\UFSCar}
	\ead{celsovb@df.ufscar.br}
	
	\author{Alan C. Santos~\orcidlink{0000-0002-6989-7958}}
	\address{\CSIC}
	\ead{ac\_santos@iff.csic.es}

	\begin{abstract}
		A fast and scalable scheme for multi-qubit resetting in superconducting quantum processors is proposed by exploiting the feasibility of frequency-tunable transmon qubits and transmon-like couplers to engineer a full programmable superconducting erasing head. We demonstrate the emergence of collective effects that lead to a decoherence-free subspace during the erasing process. The presence of such a subspace negatively impacts the device's performance and has been overlooked in other multi-qubit chips. To circumvent this issue and pave the way to the device's scalability, we employ tunable frequency couplers to identify a specific set of parameters that enables us to erase even those states within this subspace, ensuring the simultaneous multi-qubit resetting, verified here for the two-qubit case. In contrast, we show that collectivity effects can also emerge as an ingredient to speed up the erasing process. To end, we offer a proposal to build up integrated superconducting processors that can be efficiently connected to erasure heads in a scalable way.
	\end{abstract}

	\section{Introduction}

    The pursuit of quantum processing units (QPUs) with long-lived qubits resulted in the creation of qubits with coherence time exceeding one hour~\cite{Wang:21single} in a single trapped ion qubit and a few hours with nuclear spins~\cite{Zhong:15}. In the state-of-the-art superconducting quantum computing (QC), the coherence times for transmon qubits range from tens~\cite{Chu:22, Han:24} to hundreds of microseconds~\cite{Place:21} but may reach a few milliseconds with fluxonium qubits~\cite{Somoroff:23}. In addition, these achievements constitute important progress towards the advent of resilient QC, a fundamental operation in information processing emerges as a challenge: the system (re)initialization. As, in general, quantum algorithms need to be repeated thousands or millions of times to achieve fiducial statistical outcomes, fast and scalable approaches for quantum information erasure take place as a fundamental part of the computation.

	Due to the impossibility of resetting quantum information through coherent unitary operations~\cite{Kumar:00}, strategies for efficient reset have been proposed for different species of superconducting QPUs~\cite{Magnard:18, Reed:10, Geerlings:13, Egger:18, Sunada:22, Ciro_reset}, with a promising approach~\cite{Yoshioka:23} that exploits the interaction of working qubits coupled to quantum-circuit refrigerators~\cite{Tan:17}. In the context of single-qubit reset, fast reset can be done by applying coherent drives during short times to transfer the qubit information to unpopulated cavity modes~\cite{Geerlings:13}. Similarly, it can also be done by strongly coupling the transmon qubit to a low-Q resonator, where transitions from the qubit subspace to a third level of the transmon are used to speed up the energy dissipation~\cite{Reed:10, Sunada:22}. Alternatively, proposals without auxiliary energy levels and coherent fields can be used if we add an auxiliary qubit to mediate the interaction between the working qubit and the resonator~\cite{Ciro_reset}. However, to the best of our knowledge, these current approaches address only individual qubit resetting, requiring an individual dedicated resetting system for each working qubit, increasing the number of auxiliary components.
	
	In this work, we address the main problem of proposing a scalable and experimentally feasible erase head that is able to simultaneously and selectively dissipate quantum information in QPUs with more than one qubit. To this end, we identify and address collective effects where the critical challenge is to overcome resilient collective states against standard interactions of superconducting qubits. Such states are present in any model that has two or more qubits interacting with another shared system, and its behavior is similar to the subradiant states in atomic systems~\cite{Gegg_2018, Albrecht_2019, Cheng_2024}. They can appear as a direct consequence of the linearity of quantum mechanical systems (superposition principle) and interference phenomena~\cite{Celso_BDS_24}, causing a many-body quantum system prepared in these collective states to be unable to exchange energy with its external environment~\cite{Carmichael992, Yelin_2022}, thus creating a decoherence-free subspace. To avoid this adverse effect, we use two independent tunable couplers to connect a two-qubit QPU to a dissipative system. Focusing on efficient selective and simultaneous reset, we carry out simulations according to three distinct fundamental tasks required by any robust device: i) Erasing efficiency comparable to fault tolerance threshold~\cite{Boixo:18}; ii) Resetting times of the order of qubit control time; and iii) Scalability and high controllability.
	
	\section{The device model}\label{The device model}
	
	%%%%%%%%%%%%%%%%%%%%%%%%%%%%%%%%%%%%%%%%%%%%%%%%%%%%%%%%%%%%%%
	%%%%%%%%%%%%%%%%%%%%%%%%%%%%%%%%%%%%%%%%%%%%%%%%%%%%%%%%%%%%%%
	\begin{figure}[t!]
		\centering
		\includegraphics[width=1.0\linewidth]{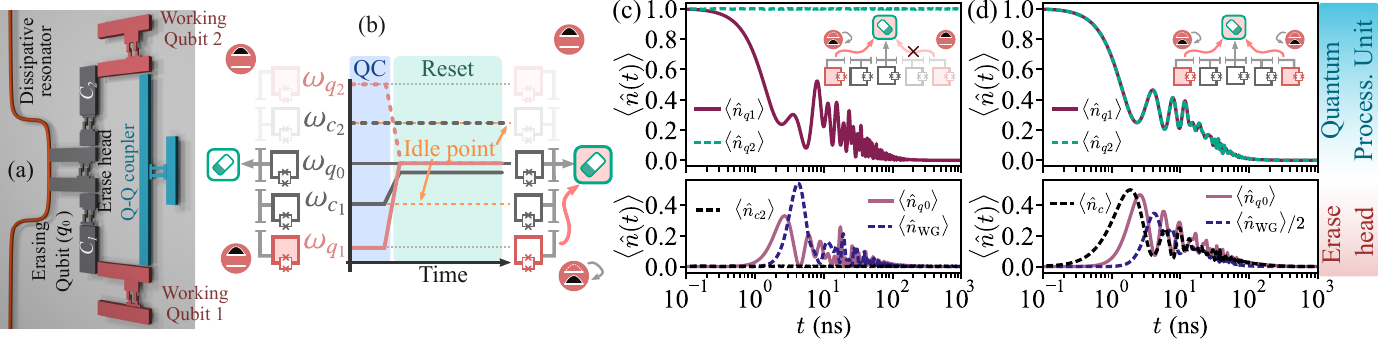}
		\caption{(a) Superconducting device proposed in this work, where two working qubits (1 and 2) interact with the erase head through couplers $C_{1}$ and $C_{2}$. A qubit-qubit (Q-Q) coupler is used during the computation stage to perform operations between the working qubits, but it is not used in the erasure stage. 
			(b) The sequence of activation of the erasing head. During the computation stage, the couplers $C_{1}$ and $C_{2}$ are at their respective idling points. After the computational stage, the frequency of the corresponding coupler of the qubit to be reinitialized is changed to the resonance with the reset frequency, while the other coupler remains at the idling point to protect its qubit from being reset. 
			(c) The population dynamics in the device during the selective process (for $q_1$), where the top graph shows the dynamics population inside the QPU (working qubits), while the bottom graphs show the average excitation number of the cavity mode, qubit $q_0$, and the coupler of the non-reset qubit. (d) shows the populations for the simultaneous resetting. 
			The parameters for the device used here are $\omega_{q_{n}} = \omega_{r} = 3 \times 2\pi$~GHz, $g_{\mathrm{qc}} = g_r = 100 \times 2\pi$~MHz, $g_{\mathrm{p}} = 3 \times 2\pi$~MHz, and $\kappa=30 \times 2\pi$~MHz. The idle and reset frequencies for the couplers are $\omega^{\mathrm{idle}}_{c_n} \approx 6.3\times 2\pi$~GHz and $\omega_{c_n} = 3.1\times 2\pi$~GHz, respectively.}
		\label{Fig:Device}
	\end{figure}
	%%%%%%%%%%%%%%%%%%%%%%%%%%%%%%%%%%%%%%%%%%%%%%%%%%%%%%%%%%%%%%
	%%%%%%%%%%%%%%%%%%%%%%%%%%%%%%%%%%%%%%%%%%%%%%%%%%%%%%%%%%%%%%
	
	The system of interest in our proposal is a superconducting quantum processor composed of transmon qubits and transmon-like couplers~\cite{Koch:PhysRevA.76.042319}. A transmon is a particular kind of superconducting qubit whose Josephson energy is much smaller than its capacitive energy. This kind of qubit has been widely used in quantum processors due to its resilience against charge noise and long coherence times~\cite{Schreier:PhysRevB.77.180502,wang2022towards,place2021new}. In this case, the Hamiltonian of a single transmon can be quantized and written as
	\begin{equation}
	H_0 = \hbar\omega_{q} \hat{a}^{\dagger}_{q}\hat{a}_{q} + \frac{\alpha}{2} \hat{a}_{q}^{\dagger}\hat{a}_{q}^{\dagger}\hat{a}_{q}\hat{a}_{q} ,
	\end{equation}
    where $\omega_{q}$ is the natural frequency transition of the qubit (ground to first excited state), and $\alpha$ is the qubit anharmonicity. In real experiments, when the contribution of the anharmonicity is large enough, compared with local driving fields, the dynamic of the system is restricted to the two lowest energy levels, $\ket{0}$ and $\ket{1}$. Therefore, the above Hamiltonian can be approximated to the qubit subspace Hamiltonian
    \begin{equation}
    	H_0 = \hbar\omega_{q} \hat{\sigma}^{+}_{q}\hat{\sigma}^{-}_{q} , \label{Eq:H0Transmon}
    \end{equation}
	where we replace the bosonic operators $\hat{a}^{\dagger}$ and $\hat{a}$ with their corresponding Pauli operators $\hat{\sigma}_{q}^{+}$ and $\hat{\sigma}_{q}^{-}$, respectively.

	As shown in Fig.~\SubFig{Fig:Device}{a}, the scalability of transmon processors can be done through the capacitive coupling between two or more transmon qubits in the same chip. This capacitive interaction between two qubits, $q_{\ell}$ and $q_{n}$, is described by the interaction Hamiltonian of the form
	\begin{equation}
		H_{\mathrm{int}} = \hbar g_{q_{\ell}q_{n}} \left(\hat{a}^{\dagger}_{q_{\ell}} + \hat{a}_{q_{\ell}}\right)\left(\hat{a}^{\dagger}_{q_{n}} + \hat{a}_{q_{n}}\right) \approx  \hbar g_{q_{\ell}q_{n}} \left(\hat{\sigma}^{+}_{q_{\ell}}\hat{\sigma}^{-}_{q_{n}} + \hat{\sigma}^{-}_{q_{\ell}}\hat{\sigma}^{+}_{q_{n}} \right), \label{Eq:Hint2}
	\end{equation}
	where the description of the Hamiltonian using Pauli matrices is allowed by high harmonicity transmons, as done in Eq.~\eqref{Eq:H0Transmon}. The final form of the Eq.~\eqref{Eq:Hint2}, which is the flip-flop excitation conserving Hamiltonian, is obtained from the Rotating Wave Approximation (RWA), as the system parameters satisfy $\omega_{q} \gg |g_{q_{\ell}q_{n}}|$. For further details on transmon qubits, we recommend the review works~\cite{Roth:23,PRXQuantum.2.040204}.

	The model presented in this work is made up of two \textit{working} qubits (used to store and process quantum information) and the full programmable erase head, as sketched in Fig.~\SubFig{Fig:Device}{a}. The erase head is composed of tunable couplers $c_{n}$ ($n=1,2$) and one qubit $q_{0}$ coupled to a dissipative resonator (or waveguide). The quits of work, $q_1$ and $q_2$ are \textit{indirectly} connected to the middle qubit $q_{0}$ through their respective couplers $c_{1}$ and $c_{2}$. Here, we assume qubit $q_{0}$ tunable in order to prevent the working qubits from the resonator dissipation. At the same time, the working qubits are also assumed tunable, although they can be frequency-fixed with frequency resonant with the cavity mode. In turn, as detailed in the next section, the couplers must be tunable to build up a highly controllable device and ensure the resetting of states within the decoherence-free subspace. Couplers and qubits are considered as two-level systems with lowering $\hat{\sigma}^{-}_{c_n}, \hat{\sigma}^{-}_{q_i}$, and raising $\hat{\sigma}^{+}_{c_n}, \hat{\sigma}^{+}_{q_i}$ operators, respectively. The lowering and raising operators for the resonator are $\hat{a}$ and $\hat{a}^\dagger$, respectively. The device Hamiltonian is given by $\hat{H} = \hat{H}_{0} + \hat{H}_{\mathrm{r}0} + \hat{H}_{\mathrm{qc}}$, with the bare Hamiltonian of the superconducting elements of the circuit (transmons, couplers and resonator) as $H_0 = \sum\nolimits_{i}\hbar\omega_{q_{i}}\hat{\sigma}_{q_{i}}^{+}\hat{\sigma}_{q_{i}}^{-} + \sum\nolimits_{n}\hbar\omega_{c_n}\hat{\sigma}_{c_n}^{+}\hat{\sigma}_{c_n}^{-} + \hbar\omega_{r}\hat{a}^{\dagger}\hat{a}$, with $\omega_{q_{i}}$, $\omega_{c_n}$ and $\omega_{r}$ the natural frequencies of the $i$-th qubit, $n$-th coupler and resonator, respectively. The qubit-resonator coupling Hamiltonian given by $\hat{H}_{\mathrm{r}0} = g_{\mathrm{r}} \hat{\sigma}_{q_{0}}^{+}\hat{a} + \mathrm{h.c.}$, with interaction strength $g_{\mathrm{r}}$, and
	\begin{equation}
		\hat{H}_{\mathrm{qc}} = \hbar\sum\nolimits_{n=1}^{2} \left[g_{\mathrm{qc}}\left(\hat{\sigma}_{q_{n}}^{+} + \hat{\sigma}_{q_{0}}^{+}\right)\hat{\sigma}_{c_{n}}^{-} + g_{\mathrm{p}}\hat{\sigma}_{q_{0}}^{+}\hat{\sigma}_{q_{n}}^{-} + \mathrm{h.c.}  \right] 
		, \label{H_full_SC}
	\end{equation} 
	describes qubits-couplers and qubit-qubit interactions, with coupling strengths $g_{\mathrm{qc}}$ and $g_{\mathrm{p}}$, respectively. For simplicity, here we assume identical qubit-couplers and qubits-qubits coupling strengths. The theory is developed by considering the chirality of the device with respect to the central qubit $q_{0}$, but the main results obtained here are achievable through non-chiral devices. The interaction of strength $g_{\mathrm{p}}$ is the capacitive parasitic qubit-qubit interaction between $q_{0}$ and the working qubits, satisfying $|g_{\mathrm{p}}| \ll |g_{\mathrm{qc}}|$~\cite{Pedro_idling_point}. To end, in our model for ideal QPUs, the resetting is done through the dissipative resonator in such a way that the system evolves under the local master equation
	\begin{equation}\label{Eq:master eq}
		\dot{\hat{\rho}}(t) = \frac{1}{i\hbar}[\hat{H}, \hat{\rho}(t)] + \frac{\kappa}{2}\left(2 \hat{a}\hat{\rho}(t) \hat{a}^{\dagger} - \{\hat{a}^{\dagger}\hat{a},\hat{\rho}(t)\}\right), \label{Eq:Mastereq}
	\end{equation}
	where $\hat{\rho}(t)$ is the density matrix of the whole system, and $\kappa$ is the effective decay rate of the resonator. Through the solution of this equation, the reset efficiency is related to the excitation number of each component of the device, given by $\langle \hat{n}_{k} \rangle = \mathrm{tr}(\hat{\rho}\hat{n}_{k})$, with $\hat{n}_{k} = \hat{\sigma}^{+}_{k}\hat{\sigma}^{-}_{k}$ for two-level systems and $\hat{n}_{\mathrm{r}} = \hat{a}^{\dagger}\hat{a}$ for the resonator. When $\langle \hat{n}_{q_n} \rangle \approx 0.0$, the $n$-th qubit is efficiently reset. 
	
	By considering experimentally feasible values for the parameters of the system~\cite{Pedro_idling_point, Houck:08, Barends:13, Bronn2015, Hutchings2017, Burnett2019, Hu:23}, we solve the above equation using numerical integration with Quantum Toolbox in Python (QuTiP)~\cite{JOHANSSON2012, JOHANSSON2013}. From the solution of Eq.~\eqref{Eq:Mastereq} expected values are computed considering the entire Hilbert space, where we trucated the Hilbert space for the bosonic mode $\hat{a}$ with $D_{\mathrm{res}}$ Fock states. $D_{\mathrm{res}} > N_{\mathrm{exc}}$ for all dynamics considered in this wotk, with $N_{\mathrm{exc}}$ the total number of excitations initially stored in the QPU.
	
	\section{Results and discussion}\label{Results and discussion}
	
	Now, we will present the main results of our work and the discussion behind the efficiency and scalability of our device. 
	
	\subsection{Selective and simultaneous resetting}
	
	As a starting point, let us show how to implement the selective and switchable delete of a single qubit in a two-qubit device, addressing the high controllability of the device. It is done in the dispersive regime of interaction, where the couplers frequency $\omega_{c_{n}}$ satisfy $|\omega_{q_{n}} - \omega_{c_{n}}| \gg g_{\mathrm{qc}}$, where the effective dynamics is described by the Hamiltonian (detailed in ~\ref{Apen_Effective Hamiltonian})
	\begin{equation}\label{H_eff_I_rewrite}
		\hat{H}_{\mathrm{eff}} = g_{\mathrm{eff}}^{(1)}\hat{\sigma}_{q_1}^{+}\hat{\sigma}_{q_0}^{-}e^{i \delta_{1} t}  + g_{\mathrm{eff}}^{(2)}\hat{\sigma}_{q_2}^{+}\hat{\sigma}_{q_0}^{-}e^{i \delta_{2} t} +g_{\mathrm{r}}\hat{\sigma}^{-}_{q_0}\hat{a}^\dagger + \mathrm{h.c.},
	\end{equation}
	with $\delta_{n} = g_{\mathrm{qc}}^2/(\omega_{r} - \omega_{c_{n}})$ and the effective couplings $g_{\mathrm{eff}}^{(n)} = g_{\mathrm{p}} + g_{\mathrm{qc}}^2/\Delta_{n}$, where we assume all qubits and resonators at frequency $\omega_{q_{n}}= \omega_{r}$, but the couplers are detuned from this resonance by $\Delta_{n} = \omega_{r} - \omega_{c_{n}}$. We assume that the selectivity of the resetting can be achieved even when all qubits of the system are in resonance with each other and with the erasing head elements. The main consequence of the above result is that $g_{\mathrm{eff}}^{(1)}$ and $g_{\mathrm{eff}}^{(2)}$ are tunable independently through its associated coupler frequency, which allows us to \textit{select} the qubit that will be effectively coupled to the resetting head system. As depicted in Fig.~\SubFig{Fig:Device}{b}, by adjusting the frequency of the $n$-th coupler at its idling frequency $\omega_{c_{n}}^{\mathrm{idle}}$, determined by $\Delta_{n}^{\mathrm{idle}} = \omega_{r} - \omega_{c_{n}}^{\mathrm{idle}} = - g_{qc}^2/g_{\mathrm{p}}$, the respective effective coupling goes to zero~\cite{Pedro_idling_point,Hu:23}. Hence, we can switch on/off the interaction of a specific qubit with the erasing head, allowing us to choose which qubit to reset.
	
	Now, to apply the selective property of our approach, we consider the resetting of the qubit $q_1$ (similar results are obtained if we choose $q_2$ instead). In this example, we initialize the system with both working qubits in their respective excited state $\ket{1}$, while the rest of the system is in the ground state. Without loss of generality, we can study the operation of the system taking the excited state $\ket{1}$ as reference for reset, once for any state of the form $\ket{\psi} = a\ket{0} + b\ket{1}$, only the component $\ket{1}$ will be affected by the resetting task. As sketched in Fig.~\SubFig{Fig:Device}{b}, during the QC stage, the couplers of the erase head are set at their respective idling points, and therefore, the QPU is not affected by the erase head. Then, reset of the qubit in the QPU can be done by switching on the interaction of the erasing head with the target qubit $q_1$ by changing the frequency of the coupler $1$ to satisfy $\omega_{r} - \omega_{c_{1}}^{\mathrm{res}} \neq \Delta_{1}^{\mathrm{idle}}$. It leads to an effective triggering of the interaction between the ``selected" qubit $q_1$ and the resetting qubit, and therefore, the information stored in $q_1$ will flow to the resonator through the finite effective interaction with the erasing head, as shown in Fig.~\SubFig{Fig:Device}{c}. 
	
	In a similar way, we can describe how to efficiently use our proposal to simultaneously reset the qubits $q_{1}$ and $q_{2}$. To achieve this result, the frequency of both couplers $1$ and $2$ have to be regulated to satisfy $\omega_{r} - \omega_{c_{n}}^{\mathrm{res}} \neq \Delta_{n}^{\mathrm{idle}}$. By doing this, the information flows simultaneously from the qubits $q_{1}$ and $q_{2}$ to the resonator through the resetting qubit, as it can be seen from Fig.~\SubFig{Fig:Device}{d}. This achievement paves the way for a scalable method to reset superconducting circuits, where a single erase head can be used to dissipate the information stored in two (or more) qubits at the same time (see~\ref{Apen_Device scalability} for further discussions).
	
	\subsection{Resetting time analysis}
	
	Once the controllable aspect of our device is demonstrated, we will now discuss the reset's efficiency and speed. Through our simulations, we have observed that an unrefined optimization of the resetting time can be done through a simple and suitable choice of the coupling strengths $g_{\mathrm{qc}}$ of the couplers with the qubit $q_{0}$, the qubit-resonator coupling $g_{r}$, and the resonator dissipation rate $\kappa$. Therefore, it is timely to discuss some conditions to be considered before any approach to computing the optimal resetting time. In fact, the time scales of interest to our device are the decay time of the resonator, given by $\tau_{\mathrm{red}} \sim \kappa^{-1}$, and the exchange time between the qubit $q_{0}$ and the resonator, given by $\tau_{\mathrm{exc}} \sim |g_{r}|^{-1}$. In this part of the device, it is important to avoid a coherent backflow of information from the resonator to the qubit $q_{0}$, leading the system dynamics to a non-Markovian regime, which will negatively affect the resetting performance of the device. For example, such a coherent backflow can be observed in Figs.~\SubFig{Fig:Device}{c-d} through the oscillations of the qubits and resonator populations. Therefore, the first condition to be considered is the weak coupling regime of $q_0$ and the resonator, expressed by the relation $\abs{g_\mathrm{r}}<\kappa$. We also can state the additional condition $g_\mathrm{qc}<g_\mathrm{r}$ for the coupling strengths in order to mitigate information backflow from the qubit $q_0$ to the coupler $c_1$. Likewise, when considering the resetting of qubit $q_2$, this last condition also helps to avoid the information backflow from the qubit $q_0$ to the coupler $c_2$. From this preliminary discussion, enhancement in the resetting time is expected when  $g_\mathrm{qc}<g_\mathrm{r}<\kappa$. This optimization strategy will be used soon to deal with the efficient reset of classes of states that are resilient against deletion.
	
	Here, the question of interest is: How fast can the simultaneous reset of two qubits be compared to the single-qubit reset performance? To address this point, one defines the \textit{resetting time}, $\tau_{\mathrm{res}}$, as the minimum time spent to delete the information with efficiency above a threshold $\Fcal_{\mathrm{tsh}}$. It is desirable that $\Fcal_{\mathrm{tsh}} = 99.5\%$, an efficiency comparable to two-qubit gate efficiency required for high-performance near-term devices~\cite{Boixo:18}. From the data shown in Fig.~\ref{Fig:Device}, we determine the resetting time for single-qubit selective reset as $\tau^{\mathrm{sq}}_{\mathrm{res}} \approx 296~$ns, which aligns to recent experimental studies in this topic~\cite{Magnard:18, Sunada:22}. Also, we computed the time $\tau^{\mathrm{sim}}_{\mathrm{res}}$ required for simultaneous reset of the qubits $q_1$ and $q_2$, and we have found the value $\tau^{\mathrm{sim}}_{\mathrm{res}} \approx 0.77\tau^{\mathrm{sq}}_{\mathrm{res}}$, that is, simultaneous resetting of two qubits is slightly faster than the single qubit case. As we shall see in the next section, this reduction by a factor $0.77$ in the time for simultaneous reset is related to a $\sqrt{2}$ factor enhancement in the effective coupling rates due to collective super-radiant-like effects during the evolution, as we shall discuss in the following. It is worth mentioning that parameter optimization is possible, which may reduce even more the resetting times~\cite{Ciro_reset}. %The decrease in the resetting time for the simultaneous reset when compared to the single-qubit case can be explained through the emergence of cooperative behavior in the system along the evolution, as we shall discuss in the following.} Essa última frase estava repetindo o que foi dito na anterior. Por isso coloque um comentário aqui.

	\subsection{Collective effects}

	\begin{figure*}[t!]
		\centering
		\includegraphics[width=1.0\linewidth]{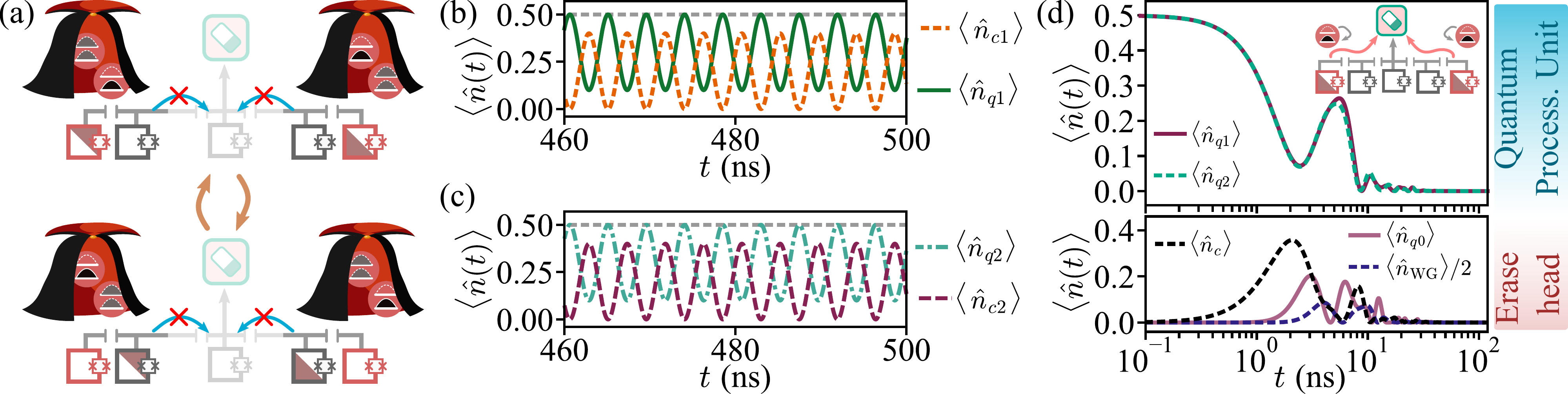}
		\caption{(a) Sketch of the trapped state dynamics of the system that inhibits the qubit reset, as the information gets trapped in the system constituted by the working qubits and their respective couplers. (b) Population dynamics for the system $q_{1}-c_{1}$ and (c) $q_{2}-c_{2}$ as function of the time. Gray dashed lines in (b) and (c) represent the total population in the system $q_{1}-c_{1}$ and $q_{2}-c_{2}$, respectively. (d) Population dynamics in the working qubits (top) and erasing head components show the simultaneous reset of a trapped state. We set $g_r = 120 \times 2 \pi$~MHz and $\kappa = 150 \times 2 \pi$~MHz, and the other parameters are as stated in Fig.~\ref{Fig:Device}.}
		\label{Fig:Collectivity}
	\end{figure*}

	\noindent Now we will study how many-body collectivity affects the device performance, in which we will pay special attention to the overlooked problem of quantum resetting devices: the emergence of collective states that would not be deleted by traditional reset approaches. To this end, let us consider the simultaneous resetting with all systems at resonance, such that the driving Hamiltonian can be written (in the interaction picture) as $\hat{H}_{\mathrm{I}} = \hat{H}_{\mathrm{qc}}$. %, for an arbitrary phase $\phi \in \Rmath$.
	Now, consider the task of deleting the two-qubit states of the form $\ket{\psi_{\phi}}_{q_{1}q_{2}} \propto \ket{10}_{q_{1}q_{2}} + e^{i\phi} \ket{01}_{q_{1}q_{2}}$, for an arbitrary phase $\phi \in \Rmath$. To this particular state, it is possible to show that for the initial state $\ket{\Psi_{\phi}} = \ket{\psi_{\phi}}_{q_{1}q_{2}}\ket{000}_{q_{0}c_{1}c_{2}}$ we get
	\begin{equation}
		\hat{H}_{\mathrm{qc}}\ket{\Psi_{\phi}} \propto g_{\mathrm{qc}} \ket{\psi_{\phi}}_{c_{1}c_{2}}\ket{000}_{q_{0,1,2}}
		+ \frac{(1 + e^{i\phi}
			) g_{\mathrm{p}}}{\sqrt{2}} \ket{1}_{q_{0}}\ket{0000}_{q_{1,2}c_{1,2}} , \label{Eq:H1psi}
	\end{equation}
	where the first term describes the initial transfer rate of information from QPU to the couplers, and the second term is the transfer rate from the QPU to the qubit $q_{0}$. As a first remark, consider the case in which $\phi = 2n\pi$, $n\in \Zmath$, where the coefficient of the second term becomes $\sqrt{2}g_{\mathrm{p}}$. It shows that if the two-qubit state has a significative projection on the super-radiant state $\ket{\psi_{\phi = 2n\pi}}_{c_1 c_2} = (\ket{1,0}_{c_1 c_2} + \ket{0,1}_{c_1 c_2})/\sqrt{2}$, during the dissipative dynamics, it leads to an enhancement in the coupling between the qubits and the erasing head through the parasitic interaction. Thus, once the information flows from the QPU to the couplers while populating the super-radiant state $\ket{\psi_\phi}_{c_1 c_2} $ for the couplers, as shown in Eq.~\eqref{Eq:H1psi}, the working qubits will have a projection on their super-radiant state during a considerable part of the dynamic. This explains why the simultaneous reset is faster than the single qubit case (See~\ref{SubAppendix} for further details). Therefore, as a first conclusion, parasitic and collateral interactions may enhance the simultaneous resetting time of states with projections on entangled states $\ket{\psi_{\phi=2n\pi}}$. This result also explains the collective enhancements of resetting Dicke states for an arbitrary number of qubits $N$~\cite{Collective_effects_qubit_reset}. In fact, we observe that such a super-radiant coupling enhancement proportional to the number of qubits $N$ may lead to an effective decrease in the resetting time of our device.
	
	Now, let us focus on the state with $\phi = (2n+1)\pi$, $n\in \Zmath$, where any transfer rate to the state $\ket{1}_{q_{0}}\ket{0000}_{q_{1,2}c_{1,2}}$ in Eq.~\eqref{Eq:H1psi} vanishes, leading to a perfectly dark state of the QPU with respect to the qubit $q_{0}$. Moreover, it is possible to show that (more details in~\ref{Apen_Effective Hamiltonian})
	\begin{equation}
		\hat{H}_{\mathrm{qc}}^{2}\ket{\Psi_{\phi = (2n+1)\pi}} \propto \hbar^2 g_{\mathrm{qc}}^2 \ket{\Psi_{\phi = (2n+1)\pi}} , \label{Eq:H2psi}
	\end{equation}
	due to destructive superposition of terms  $\hat{\sigma}_{q_{0}}^{+}\hat{\sigma}_{c_{1}}^{-}\ket{\psi_{\phi}}_{c_{1}c_{2}}\ket{000}_{q_{0,1,2}}$ and $\hat{\sigma}_{q_{0}}^{+}\hat{\sigma}_{c_{2}}^{-}\ket{\psi_{\phi}}_{c_{1}c_{2}}\ket{000}_{q_{0,1,2}}$. From this analysis one concludes that, as illustrated in Fig.~\SubFig{Fig:Collectivity}{a}, the state $\ket{\psi_{\mathrm{dark}}(0)} = \ket{\Psi_{\phi = (2n+1)\pi}}$ cannot be reset because no excitation is transferred to the resetting head, creating then a QPU-$q_{0}$ dark state. According to this result, in this case, the excitation is ``trapped" in the system composed of the couplers and working qubits. In Figs.~\SubFig{Fig:Collectivity}{b-c}, the evolution of the system shows this finding in a more concise way, which corroborates to the result suggested by the Fig.~\SubFig{Fig:Collectivity}{a}, showing the coherent population exchange between the working qubits and their respective couplers, while the population in qubit $q_{0}$ remains zero during the evolution (as the initial excitation remains ``trapped" in the systems $q_{1}-c_{1}$ and $q_{2}-c_{2}$). Under the above conditions, the presence of such trapped states leads to a decoherence-free subspace, which should also be observed in other resetting models, like the ones proposed in Refs efs.~\cite{Collective_effects_qubit_reset, Patent}, due to the collective nature of the interactions -- see~\ref{Apen_Effective Hamiltonian}. Therefore, building a device capable of handling these states is crucial to achieving robustness, which is fulfilled only through the reset of any state.

	In our model, we show below a robust way to avoid any residual amount of information stored in any initial state of the working qubits $\ket{\psi(0)}$. One possible and efficient solution can be achieved through engineering time-dependent Hamiltonians. In fact, in this way, given a trapped state for the Hamiltonian at a given time $t = 0$, such a state is affected (deleted) if it is not a trapped state of the Hamiltonian at a time $t > 0$. Also, in order to implement the protocol as fast and controllable as we can, it is desired that such a time-dependence can appear naturally during the evolution, or, in other words, without the requirement of any time-dependent external parameter. Additionally, since the set of existing trapped states depends on the device's parameters, i.e., different parameters will lead to different states not being reset (\ref{Apen_Effective Hamiltonian}), it is advantageous that the device can naturally handle these states without depending on external resources.  
	
	In order to show how our setup encompasses this property, let's first write the Hamiltonian $\hat{H}_{\mathrm{qc}}$ in the interaction picture as (with $\Delta_{c_{n}} = \omega_{r} - \omega_{c_{n}}$)
	\begin{equation}
		H_{\mathrm{I}}(t) = \hbar\sum_{n=1}^{2} \left[g_{\mathrm{qc}}\left(\hat{\sigma}_{q_{n}}^{+} + \hat{\sigma}_{q_{0}}^{+}\right)\hat{\sigma}_{c_{n}}^{-} e^{i\Delta_{c_{n}}t} + g_{\mathrm{p}}\hat{\sigma}_{q_{n}}^{+}\hat{\sigma}_{q_{0}}^{-} + \mathrm{h.c.}  \right] 
		, \label{H_full_Int}
	\end{equation}
	where we assumed that all qubits and resonator are at resonance frequency $\omega_{r}$, but the couplers are in different frequencies. Then, the key ingredients in our analysis are the time-dependent phases appearing in the first term of the Hamiltonian $H_{\mathrm{I}}(t)$. In this way, it is possible to see that when $\Delta_{c_1} \neq \Delta_{c_2}$ the state $\ket{\Psi_{\phi = (2n+1)\pi}}$ is a trapped state for the Hamiltonian only at a given time $t = 0$, but it may be affected (deleted) because it is a bright state of the evolved Hamiltonian for $t > 0$ since this state is noticed by the qubit $q_0$ and, consequently, by the dissipative mode. 
	
	As showed in Fig.~\SubFig{Fig:Collectivity}{d}, we are able to reset the state $\ket{\Psi_{\phi = (2n+1)\pi}}$ imposing the condition $\omega_{c_1} \neq \omega_{c_2}$, where $\tau^{\mathrm{sim}}_{\mathrm{res}} \approx 27$~ns. It is worth highlighting here that the suitable choice of frequency values for this case allows us to speed up the resetting procedure in comparison with Fig.~\ref{Fig:Device}, even when we deal with initial resilient states. The frequencies $\omega_{c_n}$ that allow the resetting of existing trapped states at $t=0$ and yield fast erasing depend on the coupling strengths and qubits frequencies. Because of that, such frequencies  $\omega_{c_n}$ may have different relationships according to the specifications of each device. In fact, the typical case is that the coupling strengths assume different values due to lithography imperfections. This leads to non-symmetric devices, but still, either the decoherence-free subspace (composed of different trapped states) and the frequencies $\omega_{c_n}$ that secure the resetting exist. Consequently, all the previous discussions can also be directly applied to these cases. Therefore, by employing the two tunable couplers, we achieve selectivity and high controllability of the resetting process while ensuring the device's robustness. In addition, a feasible and fiduciary control on these properties can be done simply by adjusting the coupler's frequencies (through external flux lines~\cite{Ni:21, Hu:23}), paving the way for our scalability feature.

	\subsection{Device's scalability}
	
\begin{figure}[t!]
		\centering
		\includegraphics[width=0.75\linewidth]{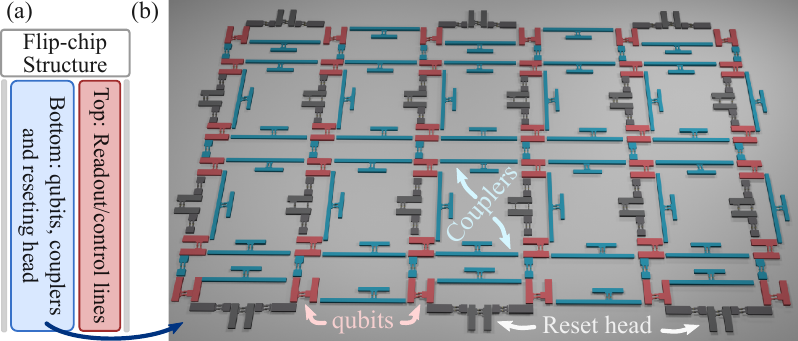}
		\caption{(a) Flip-chip structure of a (b) scalable 36-qubit QPU, with erasing heads dedicated to deleting independent cells of two qubits.}
		\label{Fig:Scalling}
	\end{figure}

	In Fig.~\ref{Fig:Scalling}, we present a pictorial way how to engineer a superconducting integrated circuit with dedicated erasing heads. One possible technology required for such a device can be found from techniques to build 3D flip-chip integrated superconducting circuits~\cite{Kosen_2022}, as depicted in Fig.~\SubFig{Fig:Scalling}{a}. Through the flip-chip approach, each plate of the chip can be fabricated separately in such a way the device consists of one \textit{processing chip}, where one deposits the qubits used for information processing, and one \textit{control chip}, where all machinery needed to control/read quantum information is placed at. Here, we propose the integration of the dedicated erasing heads in the \textit{processing chip} in such a way that the controllability of the selective and simultaneous reset can be done through additional control lines in the \textit{control chip}. Our proposal is supported by the advantages of using flip-chip processors reported in the literature concerning single-chip in planar circuits. For example, the methods for fabrication of integrated superconducting circuits developed in Ref.~\cite{Kosen_2022} allowed for a reduced degradation of the qubits performance when compared with single-chip devices, leading to enhanced gate fidelities and coherence times exceeding $90$ $\mu \mathrm{s}$ (in average). These values of coherence in 3D integrated circuits can be even longer, as those reported in Ref.~\cite{Peter:22}, where energy relaxation times around $149$ $\mu \mathrm{s}$ have been reached in a device with out-of-plane control lines~\cite{Rahamim:17}, in a similar way as sketched in Fig.~\SubFig{Fig:Scalling}{a}. In addition to these two examples, many other progresses have been done in this direction over the past couple of years~\cite{Rosenberg:17, Patterson:19, Yulin:21, Li:21, Conner:21, Gold:21}.

	By harnessing the current state-of-the-art aforementioned, we propose the $N\times M$ (for arbitrary $N$ and $M$) topology of the quantum chip. As an example, in Fig.~\SubFig{Fig:Scalling}{b}, we show a $6\times 6$ lattice with pairwise dedicated erasing heads. For the computation stage, the couplings between the working qubits are engineered through frequency tunable couplers directly connecting the two adjacent qubits, while the reset systems connect such qubits in a similar way. Through this design, the selective property of our device allows us to reinitialize any arbitrary portion of the system in a controlled and robust way. Also, it is controllable with respect to multi-qubit reset in different parts of the device simultaneously. In addition, our proposal (and its scalable example) holds a proportion of two auxiliary elements per working qubit. This is the same proportion as in simpler models, where the resetting system used to erase one qubit is composed of one resonator that is put together with a Purcell filter or coupled to an auxiliary qubit; both of them are strategies to avoid undesired dissipative effects~\cite{Reed:10, Geerlings:13, Sunada:22, Ciro_reset}.

	In an attempt to reduce this proportion, we study the resetting times if one considers adding more qubits to the original scheme of Fig.~\SubFig{Fig:Device}{a}, resulting in at least one QPU branch that contains two working qubits. Assuming three and four working qubits, the resetting times for these cases are comparable to the current qubits' lifetimes ($\tau_{\mathrm{res}} \sim 1 \mu$s). Furthermore, once we introduce more qubits to the branches, we lose part of our device's selectivity since we can no longer reset the qubits individually. Instead, we can control and reset the complete subspace connected to that branch. As a consequence, based on the current state-of-the-art, schemes like the one from Fig.~\SubFig{Fig:Scalling}{b} is the most suitable for working as quantum processors. See \ref{Apen_Device scalability} for the detailed discussion. 
	
	In~\ref{Apen_Device scalability}, we present in more detail the previous discussion for extended models, where we show how to simultaneously reset more devices with three and four working qubits using a single erasing head.

	\section{Conclusions and prospects}\label{Conclusions and prospects}
	
	In this work, we proposed a superconducting quantum eraser for fast and high-efficiency quantum information reset. The engineered device is built with a focus on its feasibility, demonstrated by its shielding performance against undesired dissipative effects and controlling the selective resetting of each qubit in a multi-qubit processor. Both of these properties are primarily achieved through the tunability of the couplers. In turn, the scalability of our model is investigated and successfully performed through the collective restart of multi-qubits devices, where we showed how the collective effects (super-radiance) may enhance the device's effectiveness. Specifically, using experimentally feasible parameters, the proposed device may achieve individual qubit erasure in approximately $300$~ns and collective erasure in $230$~ns, both with a maximum error of $0.5\%$ and without a refined parameter selection. In advance, also in this context, we identified the emergence of resilient states due to the intrinsic collective effects when a quantum eraser is coupled to more than one qubit simultaneously. However, in order to achieve a robust device, we showed how to efficiently face this neglected problem in the literature through a suitable adjustment of tunable parameters of the system. Additionally, we studied the consequences of considering more qubits in the proposed model, which did not prove advantages. Thus, we were able to design a device that controls the qubit (or qubits) to be reset independently of its state. Moreover, the task is certainly completed with high efficiency quite fast for any case, and we used the same proportion of auxiliary components as other well-known models. To end, we proposed an experimentally feasible sketch of a quantum processor composed of working qubits and qubit-qubit tunable couplers integrated with two-qubit dedicated quantum erasers. 
	
	Our work shades light on the emergence of collective phenomena as a fundamental problem for quantum information resetting and paves the way for further developments of superconducting integrated processors into scalable and robust quantum erasers.
	
	\section*{Acknowledgements}

    C M D is supported by the São Paulo Research Foundation (FAPESP, Grant No. 2022/10218-2). C J V B. acknowledges the support of the Sao Paulo Research Foundation (FAPESP, Grant No. 2022/00209-6) and the Brazilian National Council for Scientific and Technological Development (CNPq, Grant 311612/2021-0). ACS is supported by the Comunidad de Madrid through the program Talento 2024 `César Nombela', under Grant No. 2024-T1/COM-31530 (Project SWiQL). ACS acknowledges the support of the European Union's Horizon 2020 FET-Open project SuperQuLAN (899354) and the Proyecto Sinérgico CAM 2020 Y2020/TCS-6545 (NanoQuCo-CM) from the Comunidad de Madrid.

	\appendix
	
	\section{Effective Hamiltonian and robustness against resilient states}\label{Apen_Effective Hamiltonian}

	The device Hamiltonian is given by $\hat{H} = \hat{H}_{0} + \hat{H}_{\mathrm{r}0} + \hat{H}_{\mathrm{qc}}$. The bare Hamiltonian of the superconducting elements of the circuit (transmons, couplers and resonator) reads $\hat{H}_0 = \sum\nolimits^{2}_{i=0}\hbar\omega_{q_i}\hat{\sigma}_{q_i}^{+}\hat{\sigma}_{q_i}^{-} + \sum\nolimits^{2}_{n=1}\hbar\omega_{c_n}\hat{\sigma}_{c_n}^{+}\hat{\sigma}_{c_n}^{-} + \hbar\omega_{r}\hat{a}^{\dagger}\hat{a}$, where $\omega_X$ is the transition frequency of the component $X$, which can be the qubits $q_i$, the couplers $c_n$, or the cavity mode. The erase head qubit-resonator coupling Hamiltonian is $\hat{H}_{\mathrm{r}0} = g_{\mathrm{r}} \hat{\sigma}_{q_{0}}^{+}\hat{a} + \mathrm{h.c.}$, where $g_{\mathrm{r}}$ is the respective coupling strength. Also, the Hamiltonian that describes the qubit-coupler and qubit-qubit interactions is given by
	\begin{equation}\label{H_full_SC_SM1}
		\hat{H}_{\mathrm{qc}} = \hbar\sum\nolimits_{n=1}^{2} \left[g_{\mathrm{qc}}\left(\hat{\sigma}_{q_{n}}^{+} + \hat{\sigma}_{q_{0}}^{+}\right)\hat{\sigma}_{c_{n}}^{-} + g_{\mathrm{p}}\hat{\sigma}_{q_{0}}^{+}\hat{\sigma}_{q_{n}}^{-} + \mathrm{h.c.}  \right], 
	\end{equation} 
	where we assume identical qubit-coupler and qubit-qubit coupling strengths, respectively, denoted by $g_{\mathrm{qc}}$ and $g_{\mathrm{p}}$. The following derivation considers a chiral device with respect to the central qubit $q_{0}$, but the deduction for non-chiral devices is similar.

	We start moving to the interaction picture by applying the unitary transformation $\hat{U}_0(t) = \mathrm{exp}(-i\hat{H}_0t)$, with $\hat{H}_0$ defined above. This yields the transformed Hamiltoniano $\hat{H}^{(\mathrm{I})} = \hat{U}^{\dagger}_0(t)\hat{H}\hat{U}_0(t) - \hat{H}_0$. As we can rewrite the original Hamiltonian as $\hat{H} = \hat{H}_0 + \hat{V}$, where $\hat{V}= \hat{H}_{\mathrm{r0}} + \hat{H}_\mathrm{qc}$, the transformed Hamiltonian reduces to $\hat{H}^{(\mathrm{I})} = \hat{U}^{\dagger}_0(t)\hat{V}\hat{U}_0(t)$, which provides
		\begin{eqnarray}
			\hat{H}^{(\mathrm{I})} &=&  g_{\mathrm{qc}}\left(\hat{\sigma}_{q_1}^{+}+\hat{\sigma}_{q_0}^{+}\right)\hat{\sigma}_{c_1}^{-}e^{i\Delta_1 t}
			+g_{\mathrm{qc}}\left(\hat{\sigma}_{q_0}^{+}+\hat{\sigma}_{q_2}^{+}\right)\hat{\sigma}_{c_2}^{-}e^{i\Delta_2 t} +g_{\mathrm{p}}\hat{\sigma}_{q_1}^{+}\hat{\sigma}_{q_0}^{-} + \mathrm{h.c.}\nonumber \\
			&+& g_{\mathrm{p}}\hat{\sigma}_{q_0}^{+}\hat{\sigma}_{q_2}^{-} + g_{\mathrm{r}}\hat{\sigma}_{q_0}^{+}\hat{a} + \mathrm{h.c.},
		\end{eqnarray}
		where $\Delta_1=\omega_{\mathrm{r}} - \omega_{c_1}$, and $\Delta_2=\omega_{\mathrm{r}} - \omega_{c_2}$. 
	
	For detunings $\Delta_1$ and $\Delta_2$ much higher than the coupling $g_{qc}$, we can derive an effective Hamiltonian via the method of Ref.~\cite{James2, James_suprior_orders, James_suprior_orders_comment}, which is a broadly employed method for Hamiltonians that have highly oscillating terms. To apply this method, we need to find the effective contribution of the time-dependent terms that behave similarly to the time-independent ones. To this end, we first compute the effective Hamiltonian for the time-dependent terms as
		\begin{equation}\label{eff H James method}
			\tilde{\hat{H}}^{(\mathrm{I})}_\mathrm{eff} = -i\tilde{\hat{H}}^{(\mathrm{I})}(t)\int_{0}^{t}\tilde{\hat{H}}^{(\mathrm{I})}(t')dt',
		\end{equation}
		\noindent where $\tilde{\hat{H}}$ takes into account only the time-dependent terms of Eq. (A.3). The calculation of the previous equation is direct and simple since it has just exponentials. Now, with this result in hands, we add it to the time-independent terms of Eq. (A.3). Thus, by applying the rotating wave approximation (RWA) and by assuming that the couplers $C_1$ and $C_2$ remain in their respective ground states (dispersive interaction), we obtain the effective Hamiltonian
		\begin{eqnarray}
			\hat{H}^{(\mathrm{I})}_\mathrm{eff}   &=& \frac{g^{2}_{\mathrm{qc}}}{\Delta_1}\hat{\sigma}_{q_1}^{+}\hat{\sigma}_{q_1}^{-} + \left(\frac{g^{2}_{\mathrm{qc}}}{\Delta_1} + \frac{g^{2}_{\mathrm{qc}}}{\Delta_2} \right)\hat{\sigma}_{q_0}^{+}\hat{\sigma}_{q_0}^{-}  + \frac{g^{2}_{\mathrm{qc}}}{\Delta_2}\hat{\sigma}_{q_2}^{+}\hat{\sigma}_{q_2}^{-} + g_{r}\left(\hat{\sigma}_{q_0}^{+}\hat{a} + \hat{\sigma}_{q_0}^{-}\hat{a}^{\dagger}\right) \nonumber\\ &+& 
			\left(\frac{g_{\mathrm{qc}}g_{\mathrm{qc}}}{\Delta_1} + g_{\mathrm{p}} \right)\left(\hat{\sigma}_{q_1}^{+}\hat{\sigma}_{q_0}^{-} + \mathrm{h.c.}\right) + \left(\frac{g_{\mathrm{qc}}g_{\mathrm{qc}}}{\Delta_2} + g_{\mathrm{p}} \right)\left(\hat{\sigma}_{q_2}^{+}\hat{\sigma}_{q_0}^{-} + \mathrm{h.c.}\right).
	\end{eqnarray}

	Moving back to the Schr\"{o}dinger picture using the unitary transformation $\hat{H}^{\left(\mathrm{SC}\right)}_\mathrm{eff} = \hat{U}_0(t)\hat{H}\hat{U}^{\dagger}_0(t) + \hat{H}_0$, we obtain the full effective Hamiltonian
	\begin{eqnarray}\label{H_eff_sc}
		\hat{H}^{\left(\mathrm{SC}\right)}_\mathrm{eff} &=& \sum_{i=0}^{2}\Omega_{q_i}\hat{\sigma}_{q_i}^{+}\hat{\sigma}_{q_i}^{-} + \sum_{j=1}^{2}\omega_{c_j}\hat{\sigma}_{c_j}^{+}\hat{\sigma}_{c_j}^{-}  + \left(\frac{g_{\mathrm{qc}}g_{\mathrm{qc}}}{\Delta_1} + g_{\mathrm{p}} \right)\left(\hat{\sigma}_{q_1}^{+}\hat{\sigma}_{q_0}^{-} + \mathrm{h.c.}\right) \\ & +& \left(\frac{g_{\mathrm{qc}}g_{\mathrm{qc}}}{\Delta_2} + g_{\mathrm{p}} \right)\left(\hat{\sigma}_{q_2}^{+}\hat{\sigma}_{q_0}^{-} + \mathrm{h.c.}\right) + g_{\mathrm{r}}\left(\hat{\sigma}_{q_0}^{+}\hat{a} + \hat{\sigma}_{q_0}^{-}\hat{a}^{\dagger}\right)  + \omega_{r}\hat{a}^{\dagger}\hat{a},
	\end{eqnarray}
	where $\Omega_{q_1} = \omega_{\mathrm{r}} + \frac{g^{2}_{\mathrm{qc}}}{\Delta_1}$, $\Omega_{q_0} = \omega_{\mathrm{r}} + \frac{g^{2}_{\mathrm{qc}}}{\Delta_1} + \frac{g^{2}_{\mathrm{qc}}}{\Delta_2}$, and $\Omega_{q_2} = \omega_{\mathrm{r}} + \frac{g^{2}_{\mathrm{qc}}}{\Delta_2}$. Thus, from Eq. \eqref{H_eff_sc}, we can derive the effective interaction Hamiltonian in the interaction picture as
	\begin{eqnarray}\label{H_eff_I}
		\hat{H}_\mathrm{eff} &=& \left(\frac{g_{\mathrm{qc}}g_{\mathrm{qc}}}{\Delta_1} + g_{\mathrm{p}} \right)\hat{\sigma}_{q_1}^{+}\hat{\sigma}_{q_0}^{-}e^{i\left(\Omega_1 - \Omega_0\right)t}   + \left(\frac{g_{\mathrm{qc}}g_{\mathrm{qc}}}{\Delta_2} + g_{\mathrm{p}} \right)\hat{\sigma}_{q_2}^{+}\hat{\sigma}_{q_0}^{-}e^{i\left(\Omega_2 - \Omega_0\right)t}   \nonumber \\&+& g_{r}\hat{\sigma}^{-}_{q_0}\hat{a}^\dagger e^{i\left(\omega_r - \Omega_0\right)t} + \mathrm{h.c.},
	\end{eqnarray}
	which can be rewritten as (up to a global phase)
	\begin{equation}\label{H_eff_I_rewrite_SM}
		\hat{H}_{\mathrm{eff}} = g_{\mathrm{eff}}^{(1)}\hat{\sigma}_{q_1}^{+}\hat{\sigma}_{q_0}^{-}e^{i \frac{g^{2}_{\mathrm{qc}}}{\Delta_1} t}  + g_{\mathrm{eff}}^{(2)}\hat{\sigma}_{q_2}^{+}\hat{\sigma}_{q_0}^{-}e^{i \frac{g^{2}_{\mathrm{qc}}}{\Delta_2} t} +g_{\mathrm{r}}\hat{\sigma}^{-}_{q_0}\hat{a}^\dagger + \mathrm{h.c.} ,
	\end{equation}
	where $g_{\mathrm{eff}}^{(n)} = g_{\mathrm{p}} + g_{\mathrm{qc}}^2/\Delta_{n}$. Finally, for perfectly symmetric devices $\Delta_1 = \Delta_2 = \Delta$, and, consequently, $g_{\mathrm{eff}}^{(1)} = g_{\mathrm{eff}}^{(2)}=g_{\mathrm{eff}}$, it yields the effective Hamiltonian (with $\delta = g_{\mathrm{qc}}^2/\Delta$)
	\begin{equation}\label{H_eff_I_rewrite_SM2}
		\hat{H}_{\mathrm{eff}} = \Big(g_{\mathrm{eff}}\hat{\sigma}_{q_1}^{+}\hat{\sigma}_{q_0}^{-} + g_{\mathrm{eff}}\hat{\sigma}_{q_2}^{+}\hat{\sigma}_{q_0}^{-} +  g_{r}\sigma^{-}_{q_0}\hat{a}^\dagger e^{-i\delta t} + \mathrm{h.c.} \Big) .
	\end{equation}
	
	\subsection{Collective effects in simultaneous reset}\label{SubAppendix}
	
	When considering the Hamiltonian of Eq. \eqref{H_eff_I_rewrite_SM2}, we see that, due to the interaction of the dissipative cavity mode with the resetting head, the effective coupling between the erasing device and the working qubits allows the simultaneous reset for most of the states. However, a closer look will notice that the state $\ket{\tilde{\Psi}_{\phi}} = \ket{\psi_{\phi}}_{q_{1}q_{2}}\ket{0}_{q_{0}r}$, with $\ket{\psi_{\phi}}_{q_{1}q_{2}} \propto \ket{10}_{q_{1}q_{2}} + e^{i\phi} \ket{01}_{q_{1}q_{2}}$ is a dark state for the previous Hamiltonian if $\phi = (2n+1)\pi$, $n\in \Zmath$, because of the device's symmetry. To circumvent this issue, we resort to the tunability of the couplers' frequency. 
	
	Since $\Delta_1$ and $\Delta_2$ are functions of the  $\omega_{c_1}$ and $\omega_{c_2}$, respectively, they can be adjusted individually. Thus, by properly adjusting the frequency of the couplers, we can create different time dependencies on the Hamiltonian of Eq. \eqref{H_eff_I_rewrite_SM}, and, hence, the state $\ket{\tilde{\Psi}_{\phi}}$ is no longer a dark state for $t>0$. Additionally, for perfectly symmetric devices, by tuning the couplers' frequency, we change the effective coupling, which can make the dark state different from $\ket{\tilde{\Psi}_{\phi}}$. However, once the time dependency is always present for $\omega_{c_1} \neq \omega_{c_2}$, independently of the collective state in the beginning or during the dynamics, the time dependency of the Hamiltonian ensures the resetting.
	
	Now, for non-chiral devices, i.e., when we do not have perfectly symmetric devices -- the most common case, even though we have $g_{\mathrm{eff}}^{(1)} \neq g_{\mathrm{eff}}^{(2)}$ (e.g. if we have different $g_{\mathrm{qc}}$'s for the couplings between the working qubits and the couplers) for the same $\Delta$, different dark states will certainly appear. However, once again, by exploring the tunable property of the couplers' frequencies, we can create well-known time dependencies that make the existing dark states vary in time and, consequently, be reset.

	In the regime where the effective dynamics break down, i.e., when $|\Delta_n| \sim g_\mathrm{qc}$, we must consider the full original Hamiltonian to explore the potential existence of dark states, paying particular attention to the interaction part $\hat{H}_{\mathrm{qc}}$. By calculating the evolution of the state $\ket{\Psi_{\phi}} = \ket{\psi_{\phi}}_{q_{1}q_{2}}\ket{0}_{q_{0}c_{1}c_{2}r}$ through the expression $\hat{H}\ket{\Psi_{\phi}}$, it becomes clear that only the term  $\hat{H}_{\mathrm{qc}}\ket{\Psi_{\phi}}$ will be non-zero. In fact, this term gives
	\begin{eqnarray}
		\hat{H}_{\mathrm{qc}}\ket{\Psi_{\phi}} 
		&\propto&\hbar \left[
		g_{\mathrm{qc}}\ket{1}_{c_{1}}\ket{0}_{q_{1}q_{2}q_{0}c_{2}r}  
		+
		g_{\mathrm{p}}\ket{1}_{q_{0}}\ket{0}_{q_{1}q_{2}c_{1}c_{2}r}
		\right]
		\nonumber\\&+& e^{i\phi}\hbar\left[
		g_{\mathrm{qc}} \ket{1}_{c_{2}}\ket{0}_{q_{0}c_{1}q_{1}q_{2}r}
		+
		g_{\mathrm{p}}\ket{1}_{q_{0}}\ket{0}_{q_{1}q_{2}c_{1}c_{2}r}
		\right] 
		\label{SM:Eq. trapped states 1}
		\\
		&\propto&\hbar g_{\mathrm{qc}} \left[
		\ket{10}_{c_{1}c_{2}} 
		+ e^{i\phi}\ket{01}_{c_{1}c_{2}}
		\right]\ket{0}_{q_{1}q_{2}q_{0}r} 
		+ \hbar g_{\mathrm{p}}\left(1 + e^{i\phi}
		\right) \ket{1}_{q_{0}}\ket{0}_{q_{1}q_{2}c_{1}c_{2}r}. \nonumber
	\end{eqnarray}

	As previously noted in the effective dynamics, if $\phi = (2n+1)\pi$, $n\in \Zmath$, the last term in Eq. \eqref{SM:Eq. trapped states 1}, which results from the interaction between the working qubits and the resetting head, vanishes. Consequently, the evolution of $\ket{\Psi_{\phi}}$ leads to a state proportional to $\left[\ket{10}_{c_{1}c_{2}}-\ket{01}_{c_{1}c_{2}}\right]\ket{0}_{q_{1}q_{2}q_{0}r}$. This state, arising from the hopping terms in the interaction between the working qubits and their respective couplers, can be further evolved through a similar calculation, yielding
	\begin{eqnarray}\label{SM:Eq. trapped states 2}
		\hat{H}^2\ket{\Psi_{\phi}} &\propto& \hat{H}^2_{\mathrm{qc}}\ket{\Psi_{\phi=\pi}} \propto \hbar^2 g^2_{\mathrm{qc}}\left[\ket{10}_{1_{1}1_{2}}-\ket{01}_{q_{1}q_{2}}\right]\ket{0}_{c_{1}c_{2}q_{0}r} \propto \hbar^2g^2_{\mathrm{qc}}\ket{\Psi_{\phi}}.
	\end{eqnarray}
	
	Therefore, if the initial state of the system is (or evolves into) $\ket{\Psi_{\phi}}$, with $\phi = (2n+1)\pi$, $n\in \Zmath$, the population remains trapped in the working qubits and couplers, leaving the resetting head unpopulated. As a result, the resetting process will not be achieved.

	By analyzing Eqs. \eqref{SM:Eq. trapped states 1} and \eqref{SM:Eq. trapped states 2}, it is not immediately clear that adjusting the frequencies of the couplers will result in satisfactory resetting because of the lack of them in the previous equations. However, this achievement is more readily observed through the effective dynamics, and it is corroborated by the results in the main text, where we simulate the dynamics for the state $\ket{\Psi_{\phi}}$, with $\phi = (2n+1)\pi$, $n\in \Zmath$, considering the full Hamiltonian and perform the resetting successfully.

	This collective phenomenon may be observed in other models, such as those discussed in Refs.~\cite{Collective_effects_qubit_reset, Patent}, since they exhibit a similar configuration to the system studied here, which involves two systems coupled to a shared third system. However, this discussion has not been addressed before in the context of the resetting process, probably because it is challenging to identify the specific states that exhibit these novel behaviors despite concerns and the need to bypass them to achieve robust devices, that is, devices capable of resetting any state.

	\section{Device's scalability}\label{Apen_Device scalability}
	
\begin{figure}[t!]
		\includegraphics[width=1.0\linewidth]{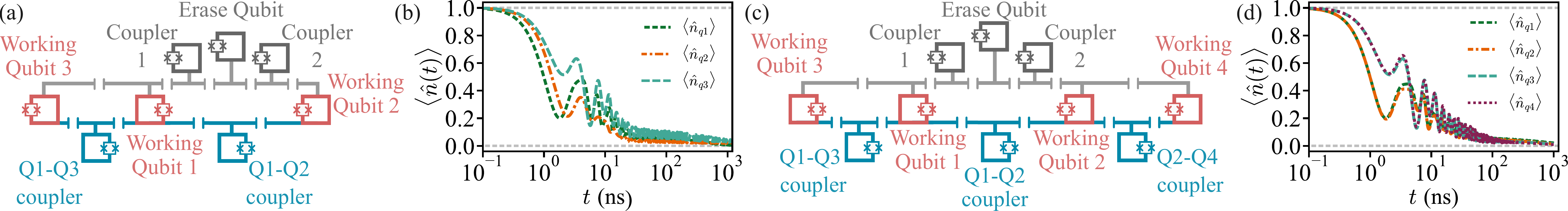}
		\caption{Panel (a) and (b) show the three-qubit system considered for the simultaneous resetting and the populations for the three working qubits case, respectively. To this case, we assume $\omega_{c_{1}} = 3.1 \times 2\pi$~GHz and $\omega_{c_{2}} = \omega_{c_{3}} = 2.9 \times 2\pi$~GHz. Panel (c) and (d) show the four-qubit system and the populations in the qubits during the simultaneous resetting, respectively, where $\omega_{c_{1}} = \omega_{c_{4}} = 3.1 \times 2\pi$~GHz and $\omega_{c_{2}} = \omega_{c_{3}} = 2.9 \times 2\pi$~GHz. The other parameters for the device used here are $\omega_{q_{i}} = \omega_{r} = 3 \times 2\pi$~GHz, $g_{\mathrm{q}} = 100 \times 2\pi$~MHz, $g_r = 120 \times 2\pi$~MHz, $g_{\mathrm{qc}}^{\mathrm{(add)}}90 \times 2\pi$~MHz,  $g_{\mathrm{p}} = 3 \times 2\pi$~MHz, and $\kappa=150 \times 2\pi$~MHz.}
		\label{Fig:SM}
	\end{figure}

	By utilizing the apparatus previously employed and considering the same resetting scheme components and configuration, additional working qubits can be incorporated into the model by coupling them to the existing qubits via couplers. Specifically, when considering the addition of one more working qubit, as depicted in Fig.~\SubFig{Fig:SM}{a}, the device Hamiltonian evolves to include the following terms
	\begin{eqnarray}
		\hat{H}_0^{(\mathrm{add})} &=& \hbar\omega_{q_3}\hat{\sigma}_{q_{3}}^{+}\hat{\sigma}_{q_{3}}^{-} + \hbar\omega_{c_3}\hat{\sigma}_{c_{3}}^{+}\hat{\sigma}_{c_{3}}^{-}, \\
		\hat{H}_{\mathrm{qc}}^{\mathrm{(add)}} &=& \hbar \left(g_{\mathrm{qc}}^{\mathrm{(add)}}\left(\hat{\sigma}_{q_{1}}^{+} + \hat{\sigma}_{q_{3}}^{+}\right)\hat{\sigma}_{c_{3}}^{-} + g_{\mathrm{p}}\hat{\sigma}_{q_{1}}^{+}\hat{\sigma}_{q_{3}}^{-} + \mathrm{h.c.}\right),
	\end{eqnarray}
	with $\hat{H}_{0}^{(\mathrm{add})}$ describing the free energies of the new components, where, following the model of Fig.~\SubFig{Fig:SM}{a}, $q_3$ represents the working qubit $3$ and $c_3$ represents the coupler $\mathrm{Q1}$-$\mathrm{Q3}$. Still, the Hamiltonian $\hat{H}_{\mathrm{qc}}^{\mathrm{(add)}}$ describes the interactions between the additional qubit and coupler, as well as the qubit-qubit interaction. Extending this framework to four working qubits, i.e., by adding two more qubits, as shown in Fig.~\SubFig{Fig:SM}{c}, the extra Hamiltonians are
	\begin{eqnarray}
		\hat{H}_0^{(\mathrm{add})} &=& \hbar\left( \omega_{q_4}\hat{\sigma}_{q_{4}}^{+}\hat{\sigma}_{q_{4}}^{-} +\omega_{q_3}\hat{\sigma}_{q_{3}}^{+}\hat{\sigma}_{q_{3}}^{-} +\omega_{c_4}\hat{\sigma}_{c_{4}}^{+}\hat{\sigma}_{c_{4}}^{-} + \omega_{c_3}\hat{\sigma}_{c_{3}}^{+}\hat{\sigma}_{c_{3}}^{-}\right);
		\\
		\hat{H}_{\mathrm{qc}}^{\mathrm{(add)}} &=&  \hbar   g_{\mathrm{qc}}^{\mathrm{(add)}}\left( \left(\hat{\sigma}_{q_{1}}^{+} + \hat{\sigma}_{q_{3}}^{+}\right)\hat{\sigma}_{c_{3}}^{-} +\left(\hat{\sigma}_{q_{2}}^{+} + \hat{\sigma}_{q_{4}}^{+}\right)\hat{\sigma}_{c_{4}}^{-}\right) 
		\nonumber \\
		&+& \hbar g_{\mathrm{p}}\left(\hat{\sigma}_{q_{1}}^{+}\hat{\sigma}_{q_{3}}^{-} + \hat{\sigma}_{q_{2}}^{+}\hat{\sigma}_{q_{4}}^{-} \right)+ \mathrm{h.c.},
	\end{eqnarray}
	where, following the model of Fig.~\SubFig{Fig:SM}{c}, $q_4$ represents the working qubit 4 and $c_4$ represents the coupler $\mathrm{Q2}$-$\mathrm{Q4}$.

	By simulating the full dynamics, we are able to reset the three and four qubits simultaneously, Figs.~\SubFig{Fig:SM}{b} and~\SubFig{Fig:SM}{d}, respectively. As shown in Fig. \ref{Fig:SM}, in both cases, the resetting times are characterized by $\tau_{\mathrm{res}} \sim 1$~$\mu$s. Even if parameter optimization can reduce the overall resetting time, these values remain comparable to the natural decay time of the working qubits. Consequently, with the current state-of-the-art superconducting qubits, incorporating five or more working qubits is not advantageous, considering the time spent during the resetting task. However, it can be more feasible shortly with the advancement of the platform and the enhancement of the qubits' lifetime.
	
	Moreover, unlike the scheme considered in the main text, the resetting process is no longer fully controllable in these expanded configurations since we cannot reset each qubit individually. For instance, in the case of three qubits, resetting qubit $q_3$ necessitates the simultaneous resetting of qubit $q_1$. Similarly, in the four-qubit case, qubit $q_2$ will be reset whenever the information stored in the qubit $q_4$ is erased. This limitation reduces both the device's control and adaptability.

	\section*{References}

	\bibliographystyle{iopart-num}
	\bibliography{bib.bib}

\end{document}